\documentclass[fleqn,usenatbib]{mnras}
\usepackage{newtxtext,newtxmath}

\usepackage[T1]{fontenc}

\DeclareRobustCommand{\VAN}[3]{#2}
\let\VANthebibliography\thebibliography
\def\thebibliography{\DeclareRobustCommand{\VAN}[3]{##3}\VANthebibliography}

\usepackage{graphicx}	
\usepackage{amsmath}	
\usepackage{tabularx}   
\usepackage{dblfloatfix}      

\newcommand{\vdag}{(v)^\dagger}
\newcommand\aastex{AAS\TeX}
\newcommand\latex{La\TeX}

\title[Selective Element Depletion in the Orion Nebula]{Self-consistent grain depletions and abundances I: The Orion Nebula as a test case}

\author[C. M. Gunasekera et al.]{
Chamani M. Gunasekera,$^{1}$\thanks{E-mail: cmgunasekera@uky.edu}
Xihan Ji,$^{1}$
Marios Chatzikos,$^{1}$
Renbin Yan$^{2}$
and Gary Ferland$^{1}$
\\
$^{1}$Department of Physics \& Astronomy, University of Kentucky, Lexington, KY 40506, USA\\
$^{2}$Department of Physics,The Chinese University of Hong Kong, Shatin, N.T., Hong Kong S.A.R., People's Republic of China\\
}

\date{Accepted \today. Received 2021 December 3; in original form 2021 September 14}

\pubyear{2021}

\begin{document}
\label{firstpage}
\pagerange{\pageref{firstpage}--\pageref{lastpage}}
\maketitle

\begin{abstract}
Atomic species in the interstellar medium (ISM) 
transition out of their gas phase mainly by depletion onto dust.
In this study, we examine if there is any change to the spectral line ratio predictions from a photoionization model of the Orion H\,{\sc ii} region when the degree of dust depletions is altered according to the most recently published model.
We use equations and parameters published by previous works, in order to streamline the calculation of depleted abundances within {\sc cloudy}. Our aim is for {\sc cloudy} users to be able to vary the level of depletion using a single parameter in the input file. This makes it possible to explore predictions for a large range of depletions more efficiently.
Finally, we discuss the results obtained for a model of the Orion Nebula when the degree of depletions are manipulated in this way. We found that the intensity of line ratios are significantly affected by depletions onto dust grains. Further, we found that adjusting dust abundances along with depletion affects the structure and the overall temperature of the H$^+$ layer across the H {\sc ii} region.
\end{abstract}




\section{Introduction}
Emission line spectra emergent from the ionized interstellar medium (hereafter ISM) are affected by the heavy-element abundances in the gas phase. Since the ISM absorbs and re-emits energy generated in stellar objects altering it from its original source \citep{spitzer-book}, the emission and absorption spectra of the ISM contain information on its physical conditions. 
Since such information obtained from emission lines is widely used by large fields of study i.e. star formation and galaxy evolution, the accuracy of the photoionization model for the ionized ISM has far-reaching consequences. 
This in turn demands an accurate representation of the gas chemical composition present in H\,{\sc ii} regions in order to ensure that model predictions are consistent with spectroscopic observations.

With the launch of \emph{Copernicus} in 1972, stellar observations of high-precision far-UV spectrometry allowed for studies on ultra-violet interstellar absorption features from heavy elements \citep{jenkins73, morton73, rogerson73a, rogerson73b, spitzer73}. Previously, element abundances in the ISM were assumed to be similar to that of cosmic abundances (also called solar abundances, because it is the chemical composition determined from absorption line measures of the Sun, stars, and meteorites normalized to 12.00 for hydrogen \citep{spitzer-book}), since it has been found that solar abundances are very similar to that in O and B stars (near which interstellar H\,{\sc ii} regions are located) \citep{salpeter77}. 
\emph{Copernicus} observations revealed that heavy element abundances are lower in the ISM than in solar abundances. The difference between the observed element composition and the solar abundances was established as being due to the depletion of gas-phase atoms onto dust grains within the ISM. These studies on \emph{Copernicus} observations also found that the degree of depletion varied from one element to the next and that the collective depletion strengths of many heavy elements varied significantly across different sightlines. Numerous studies also found a correlation between the average hydrogen density per sight line and the depletion strength \citep{savage79b, harris84, jenkins86, jenkins87, crinklaw94}. Although, the spectra of H\,{\sc ii} regions do not contain the absorption features studied in the \emph{Copernicus} observations, various other studies of emission-features have revealed the presence of dust, and the depletion of gas-phase elements onto dust-grains in these ionized regions as well \citep{spitzer-book}.

\section{Implementation details}
{\sc cloudy} is a modeling software that simulates a broad range of conditions in the interstellar matter and outputs predictions of observations, mainly spectral lines. 
It has long included depletion by grains.
The current {\sc cloudy} command that depletes elements onto grains is based on works by \cite{jenkins87} and \cite{cowie86}. The keyword {\tt\string metals deplete} without numbers on the command line, instructs the code to multiply the input abundances, by a set of scale factors specified in \emph{Hazy 1}, Table 7.8.
This depletion pattern has a few inadequacies: 
i. It does not deplete $S$ and $Ar$. 
Although debatable, there is evidence that these elements could have a considerable amount of depletion \citep{jenkins09,savage79m} Despite being a noble gas element, there is evidence that Ar "sticks" to the surface of grains, effectively depleting from the gas-phase \citep{amayo21, duley85}.
ii. The grain abundance can only be altered by the {\tt\string grains} or {\tt\string metals and grains} command, which multiplies the grain abundances by the number specified in the command line. Therefore, grain composition may not be consistent with the depletion pattern of heavy elements. 
iii. The grain mass is not consistent with the mass of depleted heavy elements since the code does not conserve mass. 
iv. There is only one built-in depletion pattern in {\sc cloudy}. While in reality, there could be a variety of depletion patterns \citep{spitzer-book,2006agna.book.....O}.

The work published in \cite{jenkins09} allows for a universal depletion pattern that is appropriate for any line of sight in an H\,{\sc i} region and can be adapted for most ISM column densities. 
Using a comprehensive survey of previously measured depletions, \cite{jenkins09} built an abstract model of the depletion pattern in the ISM, where the depletion of each element is described by a function of three simple parameters ($A_X$, $B_X$, $z_X$). Their framework also describes the collective depletion strength of the various elements using a single parameter denoted by $F_*$, which we call depletion strength. This new model of element depletion solves some of the previous problems present in the {\sc cloudy} code and provides a more up-to-date depletion pattern. Thus, the purpose of the present study is to update the {\sc cloudy} code, by incorporating an option that uses the depletion pattern as described by \cite{jenkins09}.

Note that we are using the \cite{jenkins09} model derived from the observations of the cold and warm neutral ISM, in order to study H\,{\sc ii} regions, because it is the only unified one available. It should be a high priority to build a depletion model for H\,{\sc ii} regions such as \cite{jenkins09} work did for H\,{\sc i} regions and this work is a first step in that direction.

This paper is organized as follows. Section~\ref{sec:abundance_calc} describes the new calculations adopted from \cite{jenkins09} that are being incorporated into {\sc cloudy}. We test the new depletion calculations against a model of the Orion Nebula, and a discussion of the results that follow are presented in Section~\ref{sec:orion}. Lastly, Section~\ref{sec:cloudy_mod} describes the new commands and filenames that utilizes the calculations from Section~\ref{sec:abundance_calc}. Although the new depletion framework solves many of the previous issues in the code, there are yet some caveats that need to be supplemented. For example, depletions of some important elements are missing in the new depletion pattern, there are a few elements with negative depletions which is not physical and it is not apparent how to scale grain abundances the new depletions. The fixes in the code for these shortcomings are also described in Section~\ref{sec:cloudy_mod}. 

\subsection{Calculating Post-Depletion Abundances}
\label{sec:abundance_calc}

The standard definition for gas depletion is,
\footnote{Unlike standard notations in astronomy, note that $[X/H]$ denotes log depletion of an element $X$, and 
not log abundance ratio. Abundances are denoted with curved parenthesis $(X/H)$ on a 
linear scale.}
\begin{equation}
    [X_{gas}/H] \equiv log{\{N(X)/N(H)\}}-log(X/H)_\odot,
    \label{eq:depletion_def}
\end{equation}
where $(X/H)_\odot$ are the reference abundances adopted from \cite{lodders03} and $N(X)/N(H)$ is the gas-phase abundance after depletion \citep{jenkins09}. While {\sc cloudy} provides more recent reference abundance determinations, this study uses reference abundances from \cite{lodders03} in keeping consistent with the Jenkins depletion model.
We rename the post-depleted gas-phase abundance to $(X_{gas}/H)$. 
Then solving for it we find,
\begin{equation}
    (X_{gas}/H)=(X/H)_\odot 10^{[X_{gas}/H]}.
    \label{eq:gas_dep}
\end{equation}
\cite{jenkins09} presents a generalized depletion strength $F_*$ that provides an 
empirical relationship between the individual log depletion scale factors $[X_{gas}/H]$ of different elements. This result is given in their equation 10,
\begin{equation}
    [X_{gas}/H]_{F_*}=B_X+A_X(F_*-z_X).
    \label{eq:jenkins}
\end{equation}
where $B_X$, $A_X$, and $z_X$ are parameters given in \cite{jenkins09} Table 4; the depletion parameters for the special case of S are provided in Section 9 of his paper. Parameters $B_X$, $A_X$ simply relate the value of $[X_{gas}/H]_{F_*}$ for a given element to that of other elements; so are unique to each element. As such, these depletion parameters are what determines the depletion pattern of the system. The parameter $z_X$ was introduced to make the uncertainties of $B_X$, $A_X$ independent from each other. 

\section{Impact on the star forming H\,{\sc ii} region: Orion Nebula}
\label{sec:orion}
\begin{figure*}
    \includegraphics[width=0.595\textwidth]{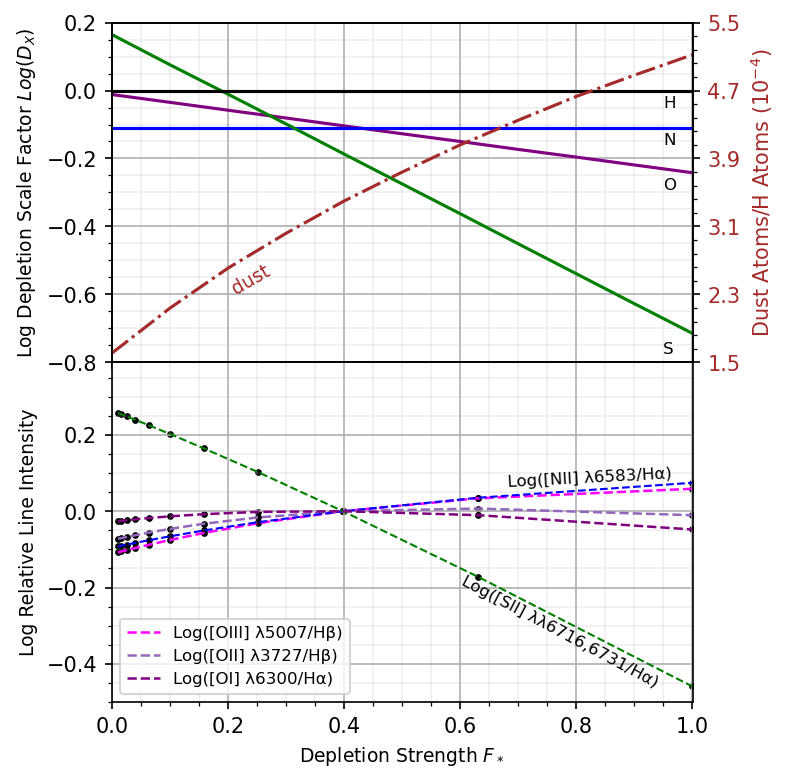}
    \centering
    \caption{Interstellar element depletion scale factors $[X/H]_{F_*}$ (top panel) and predicted line strength ratios for the Orion nebula typically found in BPT plots subtracted by its own value at $F_*=0.4$ (bottom panel) vs. depletion strength $F_*$. The red dashed line on the top panel represents the number of heavy element atoms locked in dust grains relative to the number of H atoms as a function of $F_*$. All data presented in this plot is obtained by output from {\sc cloudy}. Note that the normalization of line ratios at $F_*=0.4$ is to present the data in a way that is easily visually comparable. Furthermore, here grains are kept at their default Orion grain abundance.}
    \label{fig:orion}
\end{figure*}

The Orion Nebula is one of the nearest, brightest, and most 
observed H\,{\sc ii} regions.
The nebula's element abundances have been well studied \citep{rubin91, baldwin91, peimbert77} along with the grains. It is a blister H\,{\sc ii} region, which is an ionized layer on the surface of the background OMC1 molecular cloud. The well-understood geometry and the extensive literature makes the Orion Nebula a suitable candidate to investigate the impact of dust depletion strength on predicted emission line spectra. \cite{baldwin91} developed a photoionization model of the Orion nebula which accounts for the emission from ionized gas as well as dust grains. The input file {\tt\string orion\_hii\_open.in} specifying the physical parameters of this model can be found under the {\sc cloudy} repository {\tt\string tsuite/auto}. In Figure~\ref{fig:orion}, the top panel shows the log of the depletion scale factor of elements S, O, N, and H as a function of depletion $F_*$, and the bottom panel depicts the line ratios predicted by the Orion Nebula model. 

Even at a glance, it is apparent that the line ratios are affected to varying degrees by altering the depletion strength. 
We see that Sulfur is increasingly depleted with an increase in the value of $F_*$, much more than the other elements in Figure~\ref{fig:orion}. This same trend is mirrored by the [S\,{\sc ii}]/H$\alpha$ ratio, seen in the bottom panel of the figure. 
In contrast, although $F_*$ has no effect on the depletion strength of Nitrogen, and little effect on Oxygen compared to that of Sulfur, 
there is notable change to the line ratios [N\,{\sc ii}]/H$\alpha$, [O\,{\sc iii}]/H$\beta$, [O\,{\sc ii}]/H$\beta$, and [O\,{\sc i}]/H$\alpha$. Albeit, this change is little compared to [S\,{\sc ii}]/H$\alpha$, we point out that element abundances of Nitrogen and Oxygen are depleted whilst the intensity of their line ratios are increasing. This is a result of when the overall abundances of coolants decrease in the gas, the electron temperature rises (see Figure~\ref{fig:temp_profile} and discussion in Section~\ref{sec:scale_grains}), which strengthens the collisionally excited lines from elements that are not significantly depleted. In fact, we observe two competing effects that result in different trends of line ratios with $F_*$, depending on the slope of the depletion scale factor ($A_X$). On one hand, the depletion of heavy elements in the gas phase decreases the corresponding line strengths ratios. On the other hand, a rise in grain abundance, and a decline in coolant abundances increase the temperature, which in turn enhances the collisionally-excited line strengths. Hence, [S\,{\sc ii}]/H$\alpha$ exhibits a different trend to those of [N\,{\sc ii}]/H$\alpha$, [O\,{\sc iii}]/H$\beta$, [O\,{\sc ii}]/H$\beta$, and [O\,{\sc i}]/H$\alpha$. 

While it is interesting to observe such drastic results with [S\,{\sc ii}]/H$\alpha$, we should proceed with caution. The depletion of Sulfur suffers from contradicting results between various studies. Both \cite{calura09} and \cite{jenkins09} provide contrary findings to the more common consideration that sulfur is an element of zero depletion.
However, even the \cite{jenkins09} study labels Sulfur as a troublesome element, due to the low number of sightlines where the S depletion could be reliably determined. Since, Sulfur is an important coolant in H\,{\sc ii} regions, its level of depletion affects other collisionally excited lines. So this depletion model will benefit from future, more extensive studies on the depletion of sulfur. 

Our investigation shows that the ratios of a number of important optical emission lines are notably affected by the strength of dust depletion.
Dust depletion alters the chemical composition of the gas,
which is reflected by the line ratios of different elements.
Meanwhile, how the line ratios change with the depletion strength depends on whether the corresponding elements are heavily depleted.
Furthermore, this work should not be mis-interpreted to mean that H\,{\sc ii} regions have large distribution in $F_*$. In fact, a recent study on the abundances of various elements for nine different H\,{\sc ii} regions, showed a small variation in these abundances \citep{arellano21}. Hence it may be that real H\,{\sc ii} regions have a small variation in $F_*$. So specifying $F_*$ can become a critically useful tool to adjust the depletion pattern, since a more accurate representation of the depletions in any photoionized region can be obtained by manipulating a single parameter.

\section{Modifications to {\sc cloudy}}
\label{sec:cloudy_mod}
{\sc cloudy} \citep{cloudy17} currently provides the option to include dust depletion by using the {\tt\string metals deplete} command. We want to add an option to this command in which the user may vary the level of depletion of metals according to the depletion pattern described in \cite{jenkins09}, specified by the collective depletion strength $F_*$. This parameter is denoted by {\tt\string Fstar} in the code, and it has observed limits $0\leq F_*\leq1$. Note that it is possible that a line of sight may have depletion strength outside this range, as these are not absolute limits. They represent the lowest (at $F_*=0$) and highest (at $F_*=1$) recorded depletion strengths for neutral gas with low uncertainties.

A version of the \cite{jenkins09} Table 4, including S depletion parameters, is stored as a file called 
{\tt\string Jenkins09\_ISM\_Tab4.dep} 
in the {\tt\string data/abundances} directory. 
{\sc cloudy} utilizes this input {\tt\string.dep} file to compute the depletion scale factor $D_X$ of an element $X$ for a given $F_*$ value using,
\begin{equation}
    D_X=10^{B_X+A_X(F_*-z_X )}.
    \label{eqn:dep_scale}
\end{equation}
This factor multiplies the reference abundance of $X$, 
\begin{equation}
    (X_{gas}/H)_{F_*}=(X/H)_\odot D_X
\end{equation}
to produce the post-depleted gas-phase abundance. The complete command syntax for this modification will be described in the next major release of {\sc cloudy}. Note that this new command only depletes metals, while ISM grain abundances may be manipulated as before using the {\tt\string grains} command.

Although the \cite{jenkins09} investigation adopts their ISM reference abundances from \cite{lodders03}, the updated {\sc cloudy} distribution will provide both the \cite{lodders03} and \cite{lodders09} versions of solar abundances already converted to a linear $H=1$ scale. Users have the option of using one of the provided reference abundance files, or any other reference abundance file, by specifying it within the command, since the new depletion model is independent of the reference abundance set (because the depletion parameters in the Jenkins model were derived by analyzing differential changes in atomic gas abundances rather than studying the absolute depletions). 

\subsection{Missing Elements}
\label{sec:missing_elem}
The depletion parameters ($A_X, B_X$ and $z_X$) of Li, B, Na, Al, Ar, K, Ca are important for a complete description of a photoionization model. However, \cite{jenkins09} does not include the depletion of these elements in their depletion pattern. We supplement the depletion parameters of the above-mentioned heavy elements using data provided in \cite{savage79m}. 

The $[X_{gas}/H]$ values for the elements listed above are provided by \cite{savage79m} for two lines of sight, $\zeta$ Pup and $\zeta$ Oph. According to \cite{jenkins09} Table 2, $\zeta$ Pup has an observed $F_*=0.32$ and $\zeta$ Oph has an observed $F_*=1.05$. Following that $z_X$ is a parameter introduced simply to make the errors of $A_X$ and $B_X$ independent of each other, we set $z_X$ to zero for the new elements. Since Equation~\ref{eq:jenkins} provides a linear relation between $F_*$ and $(X_{gas}/H)$, we can backtrack our calculations from Section~\ref{sec:abundance_calc} in order to solve for the remaining two depletion parameters ($A_X$ and $B_X$). We are then left with a simple system of equations to solve for.
\begin{equation}
    A_X=([X_{gas}/H]_{\zeta Oph} - [X_{gas}/H]_{\zeta Pup})/(1.05 - 0.32),
    \label{eq:Ax}
\end{equation}
\begin{equation}
    B_X=[X_{gas}/H]_{\zeta Oph} - 1.05A_X.
    \label{eq:Bx}
\end{equation}

Note that \cite{savage79m} do not provide the depletion value of Lithium for the $\zeta$ Oph sight line ($F_*=1.05$), thus we require a different method for this element. Since lithium and silicon tend to have similar depletion values over many lines of sight \citep{savage79m},
we borrow the depletion parameter $A_{Si}$ for that of Lithium. Then, using Equation~\ref{eq:Bx} we have sufficient information to calculate $B_{Li}$. In addition to the above mentioned {\tt\string.dep} file, we include another file with the calculated parameters of the supplemental elements called {\tt\string Jenkins09\_ISM\_Gunasekera21.dep}.

\begin{table}
	\centering
	\caption{Calculated depletion parameters for missing elements from \citep{jenkins09}}
	\label{tab:missing_elements}
	\begin{tabular}{lcccc} 
		\hline\hline
		Elem. & $[X/H]_{F_*=0.32}$ & $[X/H]_{F_*=1.05}$ & $A_X$ & $B_X$ \\
		(1) & (2) & (3) & (4) & (5)\\
		\hline
        Li & -1.439 &    -   & -1.136 & -0.245 \\
        B  & -0.194 &  0.426 & -0.849 &  0.698 \\
        Na & -0.884 & -2.395 &  2.071 & -3.059 \\
        Al & -3.318 & -0.887 & -3.330 &  0.179 \\
        Ar & -0.675 & -0.298 & -0.516 & -0.133 \\
        K  & -0.999 & -0.902 & -0.133 & -0.859 \\
        Ca & -3.681 & -2.351 & -1.822 & -1.768 \\
		\hline
	\end{tabular}
	
	{NOTE -- Data for columns (2) and (3) is taken from \citep{savage79m}, Fig. 3. Column (2) corresponds to $\zeta$ Pup, and column (3) corresponds to $\zeta$ Oph. Columns (4) and (5) are calculated using Eq.~\ref{eq:Ax} and \ref{eq:Bx}, respectively. For all the above-listed elements, $z_X$ has been set to zero, since it is simply a parameter to separate the errors of $A_X$ and $B_X$.}
\end{table}

\subsection{Limits of Depletions}
\label{sec:dep_lim}
{\sc cloudy} users should make note that some $[X_{gas}/H]_{F_*}$ values for a few elements at low depletion strengths, are greater than $0$. That is, at this particular depletion strength, these elements exhibit greater gas-phase abundance than their reference abundance (shown in the top panel of Figure~\ref{fig:orion}, where the log depletion scale factor of Sulfur exceeds unity at $F_*<0.2$). 
It is physically implausible to have more atoms of an element in the gas phase after depletions.
In order to correct this error in our model, we have enabled {\sc cloudy} users to provide a maximum cut off for $[X_{gas}/H]_{F_*}$, using the keyword {\tt\string LIMIT}.

\subsection{Scaling Grains}
\label{sec:scale_grains}
When the ISM gas is depleted by a higher degree, the number of atoms in the dust phase increases. This may result in a greater number of grains with the same grain properties, or larger-sized grains \citep{spitzer-book}. Photoelectric heating by extra dust grains would then raise the equilibrium temperature, which in turn affects the strength of the collisionally excited lines. Therefore, in order to get accurate spectral line predictions for a given model, it is necessary to scale grain abundance in {\sc cloudy} along with $F_*$.

\begin{figure*}
    \includegraphics[width=\columnwidth]{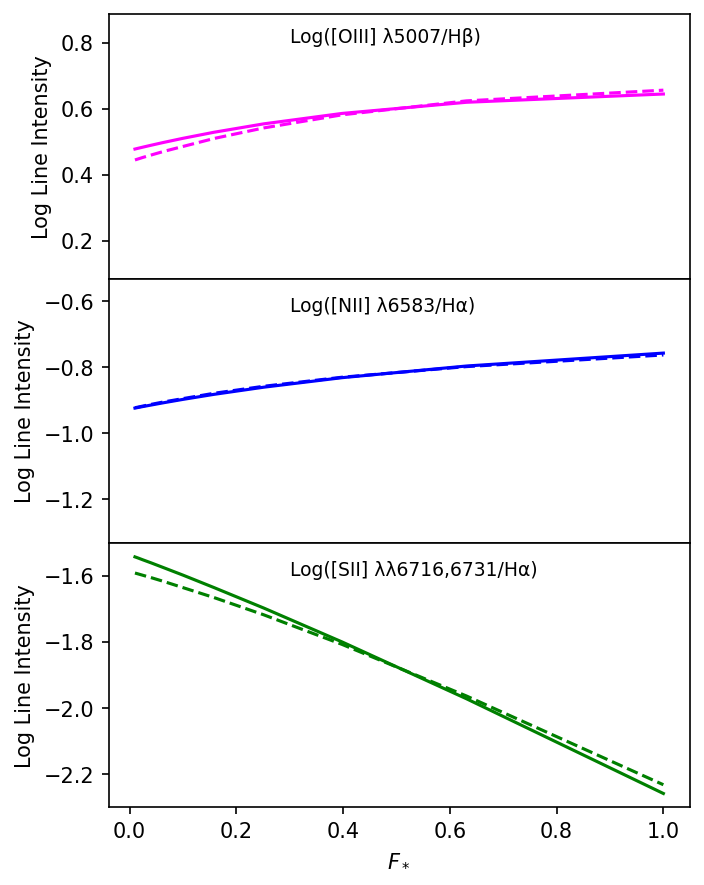}
    \includegraphics[width=\columnwidth]{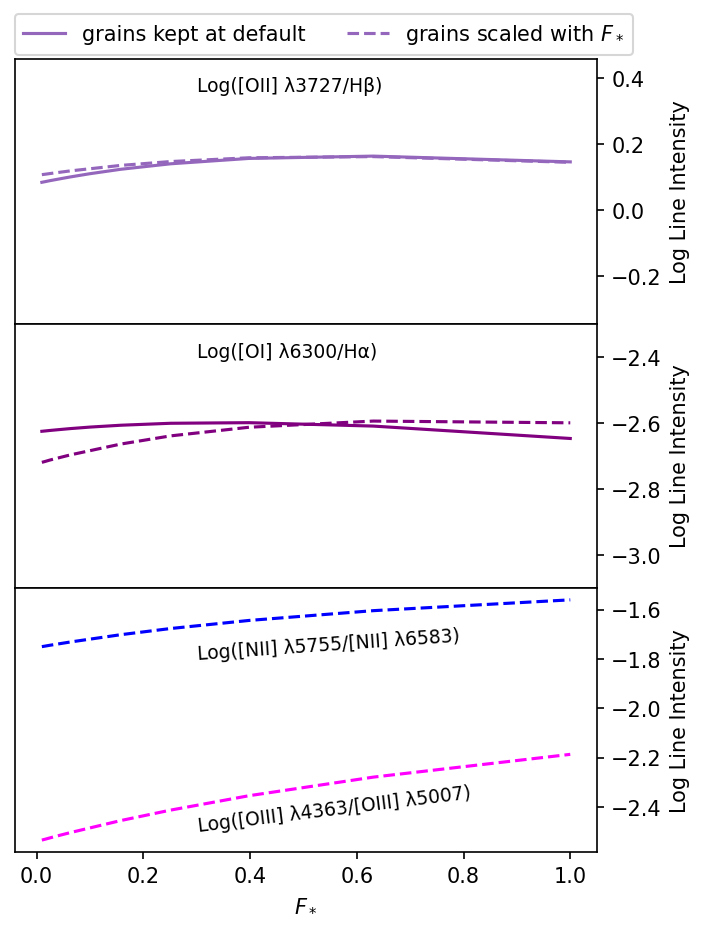}
    \caption{Predicted line strength ratios for the Orion nebula typically found in BPT plots vs. depletion strength $F_*$. All data presented in this plot is obtained by output from {\sc cloudy}. This is the same data that appears in the bottom panel of Figure~\ref{fig:orion}, except without normalizing to intensities at $F_*=0.4$.}
    \label{fig:line_grains}
\end{figure*}

Using the standard Orion grain abundance as the base value, we scaled the grain abundance for a set of $F_*$ values according to following fraction, 
\begin{equation}
fraction=\frac{\sum_X{(X_{dust}/H)_{F_*}}}{\sum_X{(X_{dust}/H)_{0.5}}}.
\end{equation}
Here we assume that $F_*=0.5$ represents the default set used by {\sc cloudy} because \cite{hensley_draine21} adopts $F_*\approx0.5$ for general star forming galaxies, and because both sets have similar depletions.
Such an analysis yields Figure~\ref{fig:line_grains}, where the solid lines represent the model results with grain abundance kept at its default {\sc cloudy} value and the dashed-style lines represent results when grains are scaled as described above (see Section \ref{sec:orion} for a more detailed discussion on the Orion nebula model in {\sc cloudy}). 
In Figure~\ref{fig:line_grains}, most line ratios exhibit little difference in intensities between the dashed and solid lines.
In addition,  most line ratios are slightly diminished when grains are scaled down at low $F_*$, and slightly enhanced when grains are scaled up at high $F_*$. Line ratios [N\,{\sc ii}]/H$\alpha$ and [O\,{\sc ii}]/H$\beta$ are exceptions to this. 
A possible explanation is that dust results in photo-electric heating of the gas, which in turn enhances forbidden lines. So when the dust is scaled down, less heating by dust causes less enhancement of line-ratios, and vice versa when the dust is scaled up for high $F_*$.
The intensity of [N\,{\sc ii}] is not changed by scaling grains since the strength of nitrogen depletion is constant at all $F_*$. Among line ratios checked, [O\,{\sc i}]/$\rm H{\alpha}$ exhibits the most change as we alter grain abundance. A discussion in the next paragraph involving temperature profiles helps explain some of this behavior. 
This change to the intensities of line ratios when grains are scaled compared to when they are not scaled is very minimal to none. This suggests that, under the assumption that grain abundance is directly proportional to the number of atoms in dust grains, scaling grains with $F_*$ does not have a significant impact on the line intensities.

\begin{figure*}
    \includegraphics[width=\textwidth]{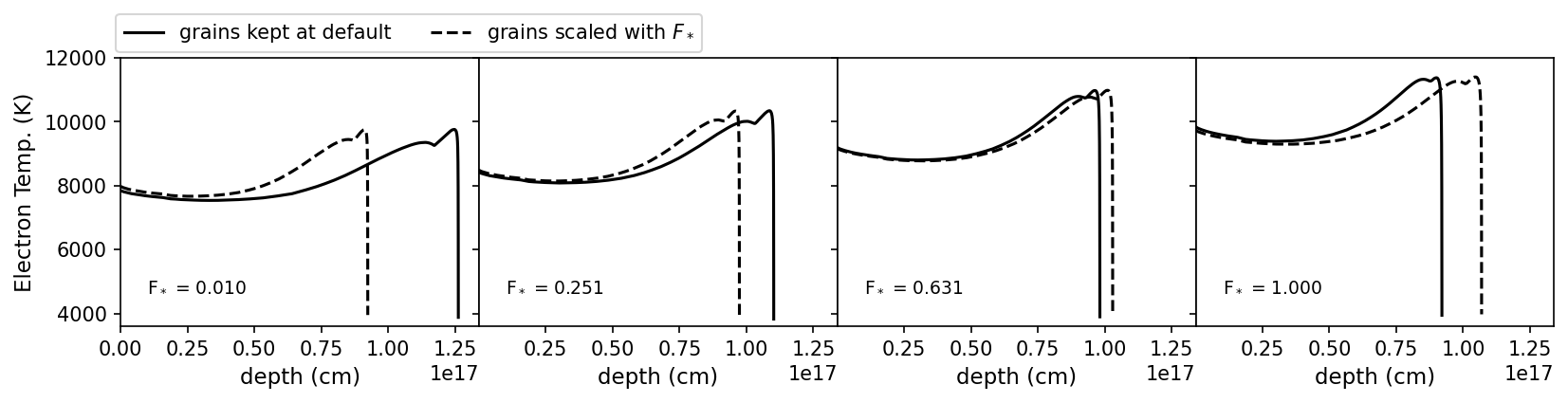}
        \centering
    \caption{Temperature profile of the H$^+$ layer across the H\,{\sc ii} region. Panels from left to right indicate temperature profile with increasing depletion strength $F_*$. The dashed line corresponds to the models where grains were scaled as in Figure~\ref{fig:line_grains}, and the solid line corresponds to grain abundance kept at {\sc cloudy}'s default Orion grain abundance.}
    \label{fig:temp_profile}
\end{figure*}

\begin{figure*}
      \includegraphics[width=0.5\textwidth]{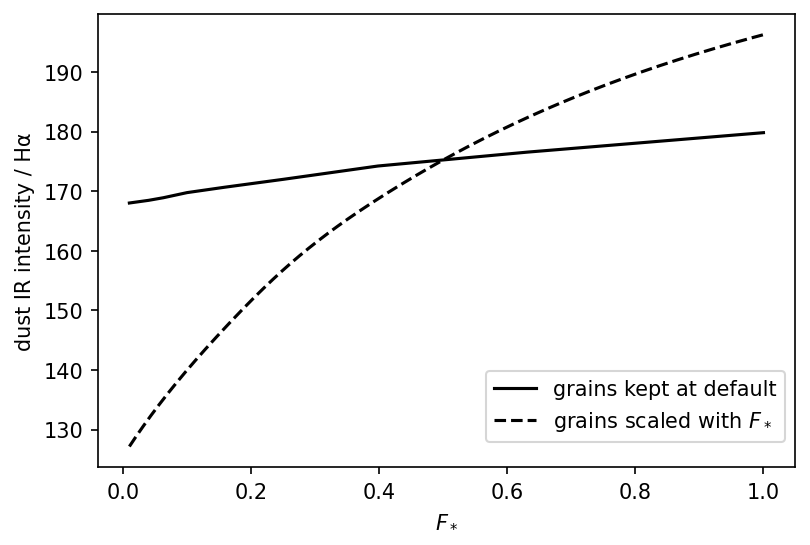}
    \centering
    \caption{IR emissions by dust in the H$^+$ layer across the H\,{\sc ii} region, as grain abundance is scaled with $F_*$ (solid line) and kept at default Orion grain abundance (dashed line), both relative to the H$\alpha$ line.}
    \label{fig:dust_ir}
\end{figure*}

Scaling grains should affect the temperature profile on an H\,{\sc ii} region, as grains contribute to heating the ISM via photoelectric emissions, and change the location of the ionization front by reducing the number of ionizing photons. 
Dust absorbs far UV radiation from starlight, then re-radiates it as IR radiation. This results in the dust heating up, as well as the photo-ejected electrons heating up the gas \citep{spitzer-book}.
So, if an increase in $F_*$ is expected to increase grain abundance, then we expect depletion strength to have a positive correlation with the overall temperature of the system.

The plots in Figure~\ref{fig:temp_profile} and the bottom-most right panel in Figure~\ref{fig:line_grains} show that $F_*$ has two effects on the temperature profile. The dashed lines in both figures correspond to the model results when grains were scaled as in Figure~\ref{fig:line_grains}, and the solid line corresponds to when grains were left at their default Orion abundance. First, we observe that Figure~\ref{fig:temp_profile} and the line ratios [O\,{\sc iii}]$\lambda$4363/[O\,{\sc iii}]$\lambda$5007 and [N\,{\sc ii}]$\lambda$5755/[N\,{\sc ii}]$\lambda$6583 (which are electron temperature tracers) agree with our expectation that increasing $F_*$ will increase the overall temperature, for both scaled and non-scaled grain cases. Increasing depletion strength increases grain abundance while depleting coolants, thereby increasing the overall temperature of the H\,{\sc ii} region.
Second, scaling grains using our method has a significant effect on the location of the H ionization front. 
Figure~\ref{fig:temp_profile} depicts that the H ionization front appears at greater depths for $F_*<0.5$ when grains are scaled, compared to when they are not. 
In addition, increasing $F_*$ for scaled grains decreases the depth of the H front. The opposite effect occurs when grains are not scaled with $F_*$, resulting in the H ionization front for scaled grains appearing at a shallower depth at $F_*>0.5$, than that for non-scaled grains. 

Notice in Figure~\ref{fig:dust_ir} when grains are scaled, the IR emission from dust increase significantly with $F_*$ relative to the emission when grains are not scaled. This demonstrates that as the dust abundance increases, it absorbs more ionizing photons, reducing the depth of the H$^+$ layer discussed in Figure~\ref{fig:temp_profile}.
The ionization structure of the cloud is thus significantly affected once the grains are scaled. 
It is expected that the size of the hydrogen partially ionized zone (a.k.a H$^+$ layer across the H\,{\sc ii} region) inside the clouds may be sensitive to the changing number of ionizing photons.
Since the line ratio [O\,{\sc i}]/H${\alpha}$ is more sensitive to the size of the partially ionized zone than the other line ratios, this may explain why the intensity of [O\,{\sc i}]/H${\alpha}$ is most affected by scaling grains as shown in Figure~\ref{fig:line_grains}.

\begin{figure*}
    \includegraphics[width=0.48\textwidth]{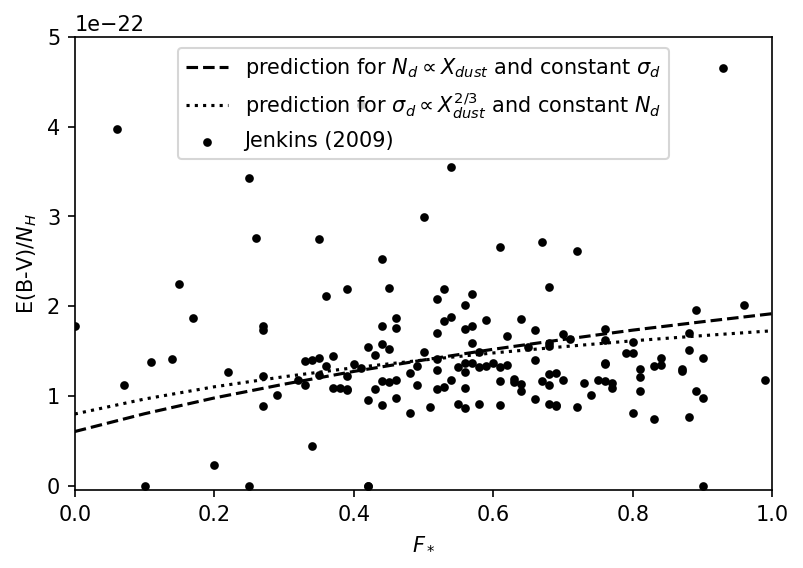}{(a)}
    \includegraphics[width=0.48\textwidth]{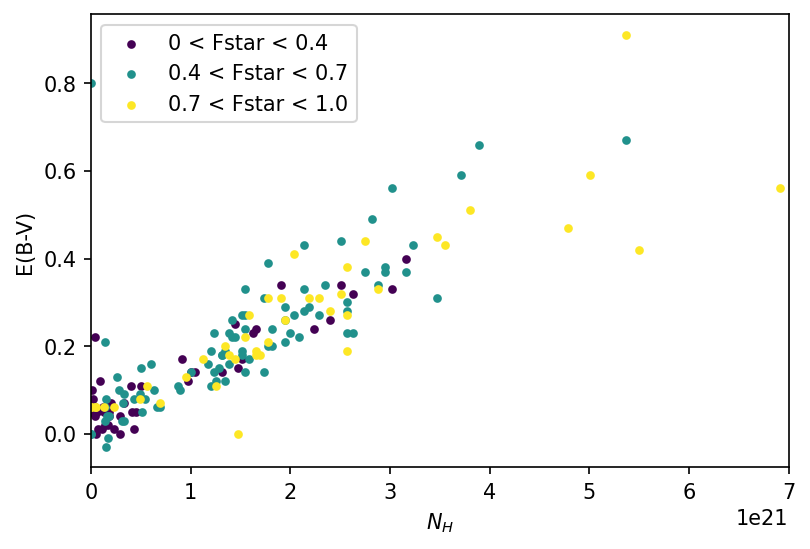}{(b)}
    \caption{(a) Depletion strength $F_*$ vs. color excess $E(B-V)$ per hydrogen column density $N_H$. (b) Color excess $E(B-V)$ vs. hydrogen column density $N_H$. Both plots show the same data. The data are obtained from the lines of sight listed in \protect\cite{jenkins09}.}
    \label{fig:scale_extinction}
\end{figure*}

The absolute extinction of spectral lines emerging from a cloud is proportional to the product of the grain column density and the geometrical cross-section of a single grain \citep[see equation (7-1) of][]{spitzer-book}. Hence the grain abundance may be alternatively found indirectly via observing the variation of the extinction of the gas with depletion strength $F_*$. Figure~\ref{fig:scale_extinction} shows two plots comparing extinction to $F_*$ for the set of sightlines listed in \cite{jenkins09}. The dashed line in Figure~\ref{fig:scale_extinction}{(a)} represents the expected trend, if only the number of grains ($N_d$) changed with $F_*$ with the size of individual grains kept constant. The dotted line represents the expected trend if only the geometrical cross-section of grains ($\sigma_d$) changed with $F_*$ with a constant number of grains. Let $\sum_X{(X_{dust}/H)_{F_*}}$ denote the total number of atoms in dust phase at a specified $F_*$ with $X_{dust,F_*}$. Given the \cite{spitzer-book} result
$A_V\propto N_d\cdot\sigma_d$ and that $N_d \propto X_{dust,F_*}$, $\sigma_d \propto (X_{dust,F_*})^{2/3}$, we expect the relations
\begin{equation}
E(B-V) = \overline{E}(B-V)\cdot\frac{X_{dust,F_*}}{X_{dust,0.5}},
\end{equation}
and
\begin{equation}
E(B-V) = \overline{E}(B-V)\cdot \left( \frac{X_{dust,F_*}}{X_{dust,0.5}} \right)^{2/3},
\end{equation}
respectively, where $\overline{E}(B-V)$ is the average color excess at $F_*=0.5$. Here we have assumed that $R_V$ does not change with $F_*$.
Figure~\ref{fig:scale_extinction}{(b)} shows the correlation between $E(B-V)$ and $N_H$ for three different bins of $F_*$ values. Both analyses show only random scatter, with no discernible correlation between $F_*$ and $E(B-V)$. Although the expected lines have a small slope, the line of sight seems to concentrate around the average color excess for all values of $F_*$. In addition, there are many sightlines with low $F_*$ but high $E(B-V)$ and many with high $F_*$ but low $E(B-V)$. Figure~\ref{fig:scale_extinction}{(b)} further establishes that there is no observable trend, as all three bins of $F_*$ values seem to populate the higher and lower ratios of $E(B-V)$ to $N_H$.
Hence at this moment, no clear relation between extinction and depletion strength can be established.

We conclude it is better not to include scaling grains with $F_*$ in the current depletion model, since: 1. Scaling grain abundance using total atoms in the dust phase shows a minimal change in spectral line ratios. 2. We could not discern a relationship between depletion strength and extinction based on the current data to constrain the scaling relation for dust grains. Consequently, the new {\sc cloudy} command depletes gas abundances only. Grain abundances can be manipulated at the user's discretion, as before using the {\tt\string grains} command, independent of the new gas-depletion command.

\section{Conclusions}
\label{sec:summary}
The primary purpose of this study was to streamline the calculations of dust depletion as described by \cite{jenkins09}, making manipulation of depletion strength in a model much simpler. 
Although there have been various advances on the study of depletions of gas abundance onto dust grains, such depletions are seldom included in investigations as it has not been incorporated into any modeling software thus far to the best of our knowledge. 
The main outcomes of this paper are as follows.
\begin{enumerate}
    \item We have integrated into {\sc cloudy}, the calculations of depleted abundances of 22 elements via the depletion scale factor $D_X$ (see Eq. \ref{eqn:dep_scale}). The depletion pattern of 7 out of the 22 elements was determined using the published post-depleted abundances obtained from \cite{savage79m}. Whereas the depletion parameters ($A_X$, $B_X$, $z_X$) of the remaining 15 elements were provided in the published work of \cite{jenkins09}. As a result of some caveats in Jenkins' work, we have included a command to limit the maximum depletion of elements to remain within physically viable standards.
    \item The predicted spectrum changes significantly with $F_*$. As depletion strength is altered, we have observed a significant change in [S\,{\sc ii}]/H$\alpha$, and comparatively little change in [N\,{\sc ii}]/H$\alpha$, [O\,{\sc iii}]/H$\beta$, [O\,{\sc ii}]/H$\beta$ as a result of two competing effects. One is the decrease of heavy elements in the gas phase which would decrease the line ratio intensity of an element. The other is the increase in temperature which would increase the collisionally-excited line strength. Depending on the value of $A_X$, the competing effects result in different trends of line-ratio versus $F_*$. This provides an explanation for [S\,{\sc ii}]/H$\alpha$ showing a different trend from [N\,{\sc ii}]/H$\alpha$, [O\,{\sc iii}]/H$\beta$, [O\,{\sc ii}]/H$\beta$, and [O\,{\sc i}]/H$\alpha$.
    \item Sulfur is a strong coolant, thus changing its depletion strength changes the intensity of other line ratios. Our results show that [S\,{\sc ii}]/H$\alpha$ is more sensitive to changes in $F_*$ than the other line ratios, making S very important. However, \cite{jenkins09} notes that determining the depletion pattern of this element is challenging and its depletion factors are more uncertain than for other elements. Furthermore, H\,{\sc ii} abundances are derived from emission lines, whereas solar abundances are from absorption lines, in which the biggest difference is the depletion of Sulfur. As such, to determine which of its results are truly compelling, the depletion of S merits further investigation.
    \item $F_*$ is negatively correlated with the depth at which the H ionization front occurs and is positively correlated to the overall temperature of ISM gas. Increasing $F_*$ increases heating by dust, as well as reduces coolant abundances, thereby heating up the ISM. 
    In addition, our analysis on scaling dust abundance with $F_*$
    reveals that only the depth at which the H ionization front occurs is altered, and not the overall temperature of the ISM. 
    When grain abundance is scaled, the intensity of IR emissions from dust is positively correlated with $F_*$.
    Hence, lower grain abundance at low $F_*$ results in less dust available to absorb ionizing photons resulting in a deeper H ionization front when grains are scaled with $F_*$. This also indicates the thickness of the hydrogen partially ionized region decreases with $F_*$. Further investigation on the correlation between dust IR emissions and a $F_*$ for a statistical sample of H {\sc ii} regions, will be beneficial to establish this result.
    \item Details of dust depletions are not yet well understood, and the new {\sc cloudy} commands will provide a better way to study the subject. As atoms of elements removed from the gas phase can only go into dust phase, dust abundance must increase when elements are depleted to greater degrees. Our attempt at scaling dust depletion showed little change to line ratio intensities. Compared to the other line ratios, [O\,{\sc i}]/H${\alpha}$ was the most affected by scaling grains, which upon investigation was likely a result of a change in the hydrogen partially ionized zone size with change in grain abundance. In addition, as the dust has a hand in the extinction process of light from distant objects, it stands to reason that $F_*$ must have some relation with $E(B-V)$. However, our analysis of color excess, gathered from the literature, has yielded no such relation. Therefore, at this moment, it is premature to include the scaling of grain abundance alongside that of the gas-phase abundance. Further study is required to determine at which scale depletion strength $F_*$ affects dust abundance.
\end{enumerate}
The results contained in this paper can be generalized to other H\,{\sc ii} regions since this method of defining dust depletion using $F_*$ is appropriate for any line of sight. 
Consequently, as altering depletion strength has an impact on the predictions for even the most widely studied H\,{\sc ii} region - the Orion Nebula, so would it on other H\,{\sc ii} regions. 
Moreover, since $F_*$ allows us to adjust the depletion pattern self-consistently using a single parameter, and 
since $F_*$ affects predicted emission line spectra, which in turn impacts studies on star formation and galaxy evolution, 
we conclude that it is critical to specify $F_*$ when utilizing the model of any H\,{\sc ii} region, regardless of the distribution of $F_*$ in H\,{\sc ii} regions.

This investigation begins to realize the impact of depletion strength $F_*$ on strong line spectra of H\,{\sc ii} regions. However, our depletion model yet has many gaps to be filled. 
Although we have shown that specifying $F_*$ affects emission line spectra in the Orion Nebula model and speculated that it should generalize to other H\,{\sc ii} regions based on \cite{jenkins09} work, we have not shown empirical evidence of this generalization. Furthermore, we also do not know the distribution of $F_*$ for H\,{\sc ii} regions. The study of a statistical sample of H\,{\sc ii} regions may be of use to establish the general effect of $F_*$ on the line ratios from these regions. 
Our main results for trends with $F_*$ stems from two competing effects on the abundance of gas available to ionize, and on the abundance of coolants. A more in-depth investigation on the effects of $F_*$ on the temperature profile, and an investigation of ionization parameter with $F_*$ may be necessary to better solidify this result. 
Such investigations will be presented in a future publication.

\section*{Acknowledgements}

We thank all the people that have made this paper possible. This includes but not limited to the published work of \cite{hensley_draine21}. 
CMG was supported by STScI (HST-AR-15018 and HST-GO-16196.003-A). 
MC acknowledges support by NSF (1910687), NASA (19-ATP19-0188), and STScI (HST-AR-14556.001-A). 
GJF acknowledges support by NSF (1816537, 1910687), NASA (ATP 17-ATP17-0141, 19-ATP19-0188), and STScI (HST-AR- 15018 and HST-GO-16196.003-A).

\textit{Software:} Python 3.8 \citep{python3}, Cloudy \citep{cloudy17}.
          
\section*{Data Availability}

The data underlying this article are available in the article and a future release of {\sc cloudy}. The extinction data analyzed in this article was obtained from the online supplementary material of \cite{jenkins09}.



\bibliographystyle{mnras}
\bibliography{cloudy_compute}







\bsp	
\label{lastpage}
\end{document}